\documentstyle[preprint,prl,aps,epsf]{revtex}
\begin{document}
\preprint{\vbox{\hbox {May 1998} \hbox{IFP-759-UNC} } }

\title{\bf Neutron Electric Dipole Moment and Spontaneous CP Breaking.}
\author{\bf Paul H. Frampton and Masayasu Harada}
\address{Department of Physics and Astronomy,}
\address{University of North Carolina, Chapel Hill, NC  27599-3255}
\maketitle
\begin{abstract}
In a model where CP is spontaneously broken, the aspon model,
it is shown by extending previous arguments about B decay and the
value of $Re(\epsilon^{'}/\epsilon)$ in the kaon system,
that the neutron electric dipole moment $D_n$ is
bounded from below by $D_n \geq 10^{-29}$e.-cm.
\end{abstract}
\newpage
The electric dipole moment {\bf D} of a neutron (NEDM)
must be proportional to its spin {\bf S} and
is hence odd under T or CP and under P. The energy
{\bf D.E} is odd under CP and even under P because
{\bf E} is odd under P and even under CP. Thus
a non-vanishing NEDM would be an explicit example
of CP violation, outside of the kaon system.

The discovery of a NEDM will be very important in
formulating the theory of CP violation where the
problem is the fact that only one (complex)
parameter, $\epsilon_K$, has been accurately measured.
The second parameter in the kaon system,
$Re(\epsilon^{'}/\epsilon)$, remains uncertain -
it is still consistent with zero - and the CP
asymmetries in B meson decays have yet to
be determined. Any further experimental measurement
of CP violation will constrain the viable theories.

The experimental limit on $D_n$ is presently
$D_n \leq 10^{-25}$ e.cm.\cite{1,2} but it is possible
\cite{3} that this limit will be pushed back
by several orders of magnitude in the forseeable
future, even a limit of $10^{-30}$ e.cm. being
conceivable. On the other hand, the prediction of the KM
mechanism within the framework of the
Standard Model is $D_n \simeq 2 \times 10^{-32}$ e.cm.
\cite{a,b,c,d} which is beyond the reach of {\it any} planned experiments.

In a model of spontaneously broken CP symmetry\cite{FK,FN}
one may simultaneously solve the strong CP problem
(without an axion) and accommodate the observed CP violation
in the kaon system. In an economical model, the Standard Model
is augmented by an additional $U(1)$ gauge group
under which all the usual three generations of quarks and leptons
are neutral, as are the established twelve gauge bosons
($\gamma, Z, W^{\pm}, g^a$) and the single Higgs doublet.
One adds a non-chiral quark doublet $(U, D)$ which carries
the new charge, and two complex Higgs scalar singlets $\chi^{\alpha}$.
These $\chi^{\alpha} (\alpha = 1,2)$ carry an equal new charge.
With this arrangement, the generalized ($4 \times 4$) quark
mass matrix has a real determinant and hence at tree-level
the value of $\bar{\Theta}$ is zero if the underlying
theory respects CP symmetry. 

Denoting the (complex) VEVs of the $\chi^{\alpha}$
by $<\chi^{\alpha}>$ and the Yukawa coupling constants
to the $i$th generation by $h^{\alpha}_i$ useful 
parameters are 
\begin{equation}
x_i = \sum_{\alpha=1}^{\alpha=2} h_i^{\alpha} <\chi^{\alpha}>  \label{X}
\end{equation}
These $x_i$ are complex, but for convenience we shall write $x$ to
denote $x = |x_i|$, the modulus, and taken to be generation independent
since the limits on $x$ are not sensitive to the generation considered.

In \cite{4}, it was shown that the predicted value of $\bar{\Theta}$
is estimated from a one-loop diagram as:
\begin{equation}
\bar{\Theta} = \frac{\lambda x^2}{16 \pi^2} \label{theta}
\end{equation}
This follows from the diagram shown in Fig1. In the notation 
of \cite{FN} the imaginary part gives a contribution:
\begin{equation}
\bar{\Theta} (up) = \frac{1}{\sqrt{2}} \frac{1}{(4 \pi)^2}
\sum_{\alpha=1,l =1}^{\alpha=2,l=3} h_{l}^{\alpha} Im[x_l^{*}] \lambda_{\alpha} 
\frac{\kappa}{M}
\end{equation}
which gives the estimate of Eq.(\ref{theta}), from which it follows 
that $\lambda x^2$ is less than $\sim 10^{-8}$.
Here $\lambda_{\alpha}$ is the coefficient of the quartic interaction 
between the two types of
Higgs $|\phi|^2 |\chi_{\alpha}|^2$, and $\lambda$ with no subscript
is an average value. (Actually there are three independent $\lambda$
corresponding to indices 11, 12+21, 22 but our estimates will not
distinguish these).

The neutron electric dipole $D_n$ has been calculated in terms of
$\bar{\Theta}$ long ago\cite{5,6} with the result that
\begin{equation}
D_n \simeq 10^{-15} \bar{\Theta} e. cm. \label{Dn}
\end{equation}
and so we know from $D_n \leq 10^{-25}$ e. cm. empirically
that $\bar{\Theta} \leq 10^{-10}$. Here we seek to
establish a lower limit on the prediction for $\bar{\Theta}$ 
in the present model.

Firstly, we look at the kaon system for which the parameter
$|\epsilon_K|$ is given by\cite{FN,FH}:
\begin{equation}
|\epsilon_K| = \frac{1}{\sqrt{2} \Delta m_K} m_K \frac{f_K^2}{3} \frac{2}{\kappa^2} x^4
\end{equation}
Using $(\Delta m_K/m_K) = 7.0 \times 10^{-15}$, $f_K = 0.16$GeV gives
the relationship between $x^2$ and the $U(1)_{new}$
breaking scale:
\begin{equation}
\kappa/x^2 = 2.9 \times 10^7 GeV.
\end{equation}
Thus, if we insist that the $U(1)_{new}$ is broken above
the electroweak breaking scale ($\sim 250GeV$) then:
\begin{equation}
x^2 \gtrsim 10^{-5}.
\end{equation}
From Eq.(\ref{theta}), this means that $\lambda < 10^{-3}$.

In \cite{4}, it was further argued on the basis of naturalness that
$\lambda > 10^{-5}$ which (if taken at face value)
would imply that $\bar{\Theta} > 10^{-12}$
and hence $D_n > 10^{-27}$ e. cm. 

But here we use what is a more solid, and 
more conservative, bound to achieve
our lower limit on $D_n$. 
Our bound follows from the fact that the $|\phi|^2|\chi_{\alpha}|^2$ interaction
receives a one-loop correction
from the quark loop box where three sides are the top
quark and the fourth is the heavy $U$ quark (see Fig. 2).
The full $\lambda$ is given by:
\begin{equation}
\lambda = \lambda_{tree(bare)} + \lambda_{1-loop} (\mbox{including counterterm}) + \mbox{higher loops}.
\end{equation}
and the 1-loop finite contribution, for the dominant diagram
(Fig 2), neglecting the quark masses and taking
$h_3^{\alpha}, g_t$ as the respective Yukawa couplings
to $\chi_{\alpha}$ and $\phi$ of the third generation
\begin{equation}
\lambda_{1-loop} \simeq \int \frac{d^4k}{(2 \pi)^4} |h_3^{\alpha}|^2 
|g_t|^2 \frac{1}{k^4} + \mbox{counterterm}.
\end{equation}
which implies that
the lowest value for $\lambda$ (without accidental cancellations)
is:
\begin{equation}
\lambda \gtrsim \frac{x^2}{16 \pi^2}  \label{lambda}
\end{equation}
Combining Eqs(\ref{theta}) and (\ref{lambda}) then gives the estimate
for $\bar{\Theta}$ of 
\begin{equation}
\bar{\Theta} \geq \frac{x^4}{16 \pi^4}      
\end{equation}
which implies that $x^2 \leq 10^{-3}$ (incidentally in full 
agreement with \cite{4}) and that $10^{-10} \geq \bar{\Theta}
\geq 10^{-14}$.

From Eq(\ref{Dn}), this then give our required lower limit
on the neutron electric dipole moment of
\begin{equation}
D_n \geq 10^{-29} e. cm.
\end{equation}
This is more than two orders of magnitude greater than the 
prediction of the KM mechanism and thus provides yet another 
distinguishing feature of spontaneous CP violation.
Recall that this spontaneous CP violation model predicts,
as reflects its superweak nature, that
$Re(\epsilon^{'}/\epsilon) \simeq 10^{-5}$\cite{FH}
(compared to $O(10^{-3})$
in the KM mechanism) and that the angles
$\beta,\gamma$ of the
unitarity triangle measurable in B Factories are a few milliradians\cite{4} 
(rather than $O(1)$ as in the KM mechanism). 
It is very exciting to realize that
at least one of these different CP breakdown
mechanisms must be experimentally refuted with the advent of B Factories. 

\bigskip
\bigskip
\bigskip

This work was supported in part by the
US Department of energy under Grant No. 
DE-FG05-85ER-40219.

\vspace{20mm}

{\bf Figure Captions.}

\noindent Fig 1. One loop diagram of which the imaginary part
contributes to $\bar{\Theta}$.

\noindent Fig. 2. One loop $|\phi|^2|\chi|^2$ counterterm
for the coupling $\lambda$.

\newpage

\begin{figure}

\begin{center}

\epsfxsize=4.0in
\ \epsfbox{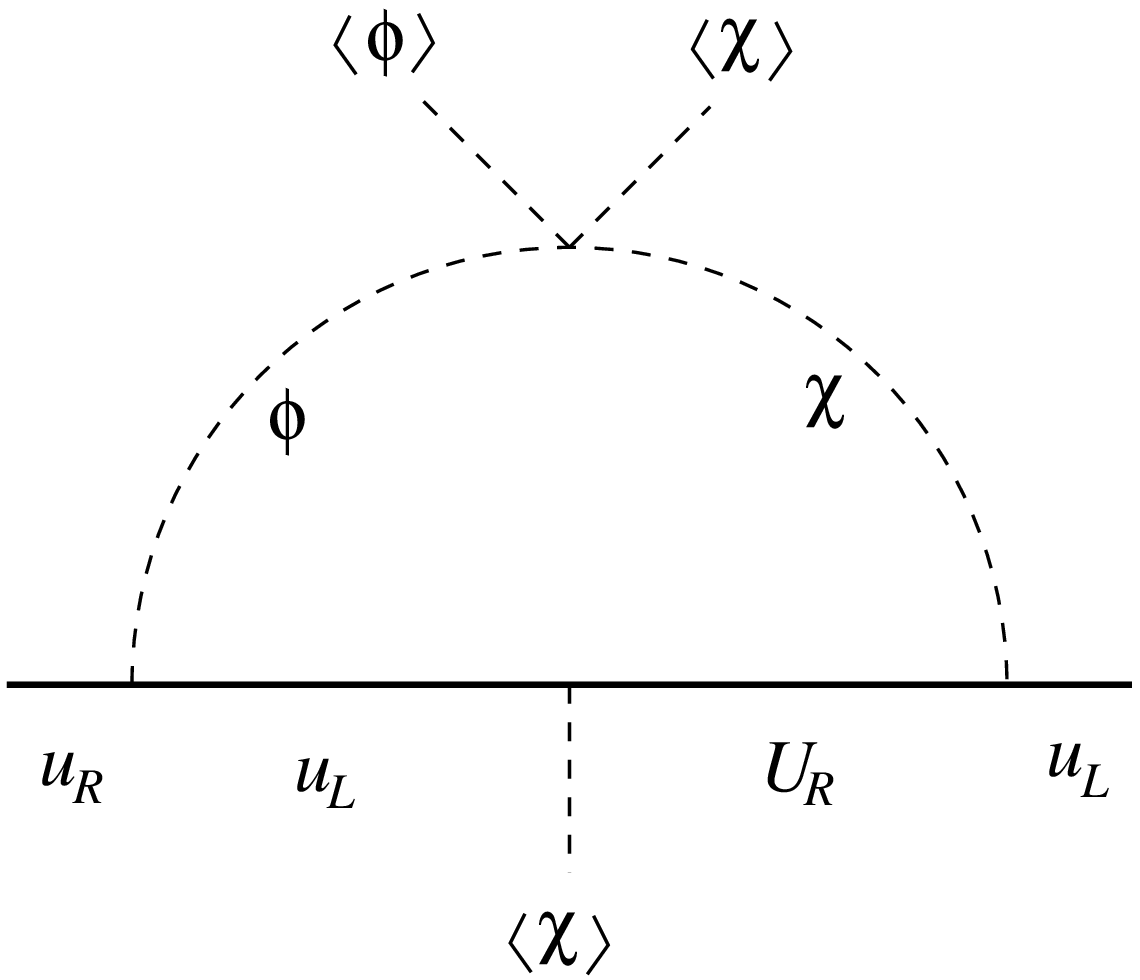}

Figure 1.
\end{center}

\end{figure}
\newpage

\begin{figure}

\begin{center}

\epsfxsize=4.0in
\ \epsfbox{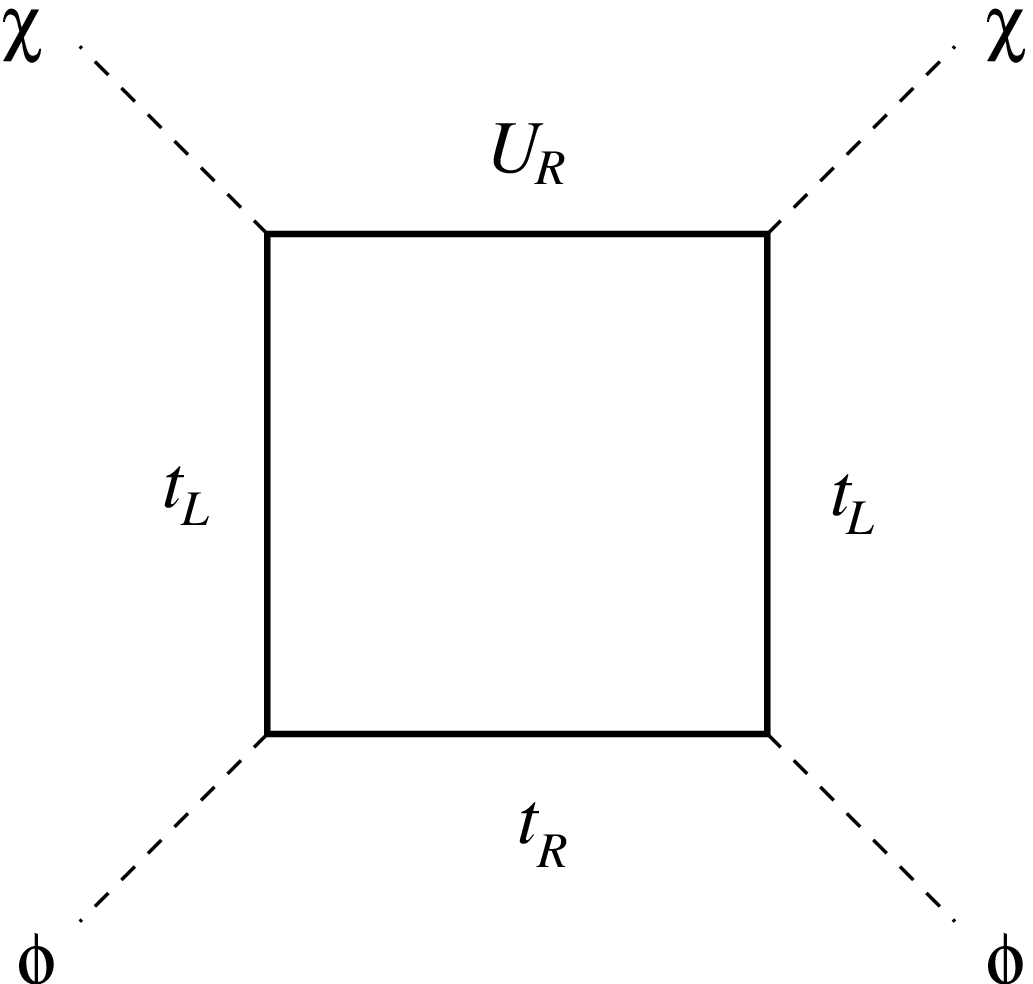}

Figure 2
\end{center}
\end{figure}

\end{document}